\documentclass{iopart}

\usepackage{graphicx}
\usepackage{mathrsfs}
\usepackage{hyperref}
\usepackage{geometry}
\usepackage{xcolor}
\newcommand{\text}[1]{\mathrm{#1}}

\begin{document}

\title{Skyrmion Transport and Annihilation in Funnel Geometries}
\author{
            F. S. Rocha$^1$,
            J. C. Bellizotti Souza$^1$, 
            N. P. Vizarim$^2$, 
            C. J. O. Reichhardt$^3$, 
            C. Reichhardt$^3$
            and P. A. Venegas$^2$}
\address{$^1$ POSMAT - Programa de P\'os-Gradua\c{c}\~ao em Ci\^encia e Tecnologia de Materiais, Faculdade de Ci\^encias, Universidade Estadual Paulista - UNESP, Bauru, SP, CP 473, 17033-360, Brazil}
\address{$^2$ Departamento de F\'isica, Faculdade de Ci\^encias, Unesp-Universidade Estadual Paulista, CP 473, 17033-360 Bauru, SP, Brazil}
\address{$^3$ Theoretical Division and Center for Nonlinear Studies, Los Alamos National Laboratory, Los Alamos, New Mexico 87545, USA}

\ead{nicolas.vizarim@unesp.br}

\begin{abstract}
Using atomistic simulations, we have investigated the transport and annihilation of skyrmions interacting with a funnel array under a current applied perpendicular to the funnel axis. We find that transport without annihilation is possible at low currents, when the motion is dominated by skyrmion-skyrmion interactions and skyrmions push each other through the funnel opening. Skyrmion annihilation occurs for higher currents when skyrmions in the upper half of the sample exert pressure on skyrmions in the bottom half of the sample due to the external current. Upon interacting with the funnel wall, the skyrmions undergo a size reduction that makes it easier for them to pass through the funnel opening. We find five phases as a function of the applied current and the size of the funnel opening: (i) pinned, (ii) transport without annihilation, (iii) transport with annihilation, (iv) complete annihilation, and (v) a reentrant pinning phase that only occurs for very narrow openings. Our findings provide insight into how to control skyrmion transport using funnel arrays by delineating regimes in which transport of skyrmions is possible as well as the conditions under which annihilation occurs.
\end{abstract}

\noindent{\it Keywords\/}: Skyrmion, Transport, Information carrier, Hall angle, Annihilation

\maketitle

\section{Introduction}
Magnetic skyrmions are topologically protected
particle-like spin configurations that have
received growing attention in recent years. 
Due to their reduced size, stability, and ease of transport,
skyrmions are considered to be a particularly promising
candidate for creating new logical and memory spintronics
devices
\cite{nagaosa_topological_2013,everschor-sitte_perspective_2018,fert_magnetic_2017}.
Skyrmions arise in chiral magnetic materials
\cite{nagaosa_topological_2013,muhlbauer_skyrmion_2009,yu_real-space_2010}
including MnSi, FeCoSi, and other B20 transition metal compounds 
\cite{muhlbauer_skyrmion_2009,munzer_skyrmion_2010,pfleiderer_skyrmion_2010, yu_real-space_2010,yu_near_2011},
and their formation is the result of an
interplay between the Dzyaloshinsky–Moriya (DM) interaction
\cite{dzyaloshinsky_thermodynamic_1958,moriya_anisotropic_1960} and ferromagnetic exchange. 

One of the most important features of skyrmions is that they can be set into motion by the application of a transport current. It was shown that the current density needed to drive skyrmions is much lower than the currents necessary to move magnetic domain walls \cite{schulz_emergent_2012}.
Thus, understanding skyrmion stability and behavior under the influence of transport currents is crucial for
the creation of future technological devices.
Due to their topology, skyrmions exhibit strong gyrotropic effects that
strongly affect their dynamics
\cite{nagaosa_topological_2013,Reichhardt22a,jiang_direct_2017}.
In particular, under an applied
spin current the skyrmions are deflected
by the skyrmion Hall effect (SHE), which
produces a finite skyrmion Hall angle $\theta_{\rm sk}^{\rm int}$. 
The magnitude of $\theta_{\rm sk}^{\rm int}$
is determined by 
the material parameters along with intrinsic defects in the sample.
In general, $\theta_{\rm sk}^{\rm int}$ falls in the range
$0^\circ$ to $55^{\circ}$
\cite{jiang_direct_2017,Reichhardt15,litzius_skyrmion_2017,zeissler_diameter-independent_2020},
but larger values of $\theta_{\rm sk}^{\rm int}$ are possible depending
on the parameters of the system.
For certain conditions that are
related to non-adiabatic spin transfer torque,
the Hall angle can be zero \cite{zhang_magnetic_2017}.
From a technological application point of view,
a finite skyrmion Hall angle is problematic because it
can cause the skyrmion to deviate towards
the sample edge and annihilate, destroying any information that may have
been associated with the skyrmion location.
There have been various efforts to understand how to control the
skyrmion motion and mitigate the intrinsic Hall angle,
including through the use of
periodic pinning
\cite{reichhardt_quantized_2015,Ma16,vizarim_topological_2020,feilhauer_controlled_2020},
ratchet effects
\cite{gobel_skyrmion_2021,reichhardt_magnus-induced_2015,Yamaguchi20,souza_skyrmion_2021}, 
1D potential wells
\cite{purnama_guided_2015,Reichhardt16a,GonzalezGomez19,Juge21,DelValle22},
nanotracks \cite{zhang_magnetic_2015,toscano_suppression_2020},
soliton motion \cite{vizarim_soliton_2022},
interface guided motion \cite{vizarim_guided_2021}
and gradients in the strain, temperature, or magnetic field 
\cite{yanes_skyrmion_2019,zhang_manipulation_2018,everschor_rotating_2012,kong_dynamics_2013}.

Several proposals for skyrmion based devices involve using a skyrmion
as the information carrier and controlling the flux of the skyrmions.
In studies of skyrmions and racetrack memories, Zhang \textit{et. al.}
\cite{zhang_skyrmion-skyrmion_2015}
not only
showed the influence and importance of skyrmion-skyrmion interactions
and the skyrmion-edge repulsion in racetracks,
but also investigated the effect of spacing between consecutive
skyrmionic bits and demonstrated
the feasibility of writing, reading, and deleting
skyrmions in future skyrmionic devices.
Tomasello \textit{et. al.} \cite{tomasello_strategy_2015}
analyzed scenarios for controlling Ne{\' e}l and Bloch
skyrmions by SHE or spin-transfer torque (STT)
and concluded that Ne{\' e}l skyrmions driven by SHE
are more promising from a technological point of view since
the skyrmion motion is less sensitive to
the defects and edge roughness that are present in real skyrmion devices.
These results pave the way for applications using skyrmions,
such as a skyrmion diode, where skyrmions flow easily in one direction
but not in the other.
A typical 
method for reducing skyrmion flow in a given direction
is to introduce a geometric constriction
or obstacles for the skyrmions.
Feng \textit{et. al.} \cite{Feng22} proposed a stripe-shaped device with lateral asymmetry that allows the skyrmion to flow in the easy direction of
the asymmetry but not in the hard direction.
Skyrmions flowing along the easy direction
follow the edge of the sample and can be transported through the device;
however, a skyrmion that flows in the hard direction
interacts with the lateral asymmetry and is annihilated.
Using a similar idea, Jung \textit{et. al.} \cite{Jung21}
proposed a skyrmion diode involving
asymmetrically shaped geometric structures. In this case,
a skyrmion moving in the hard direction
is not annihilated but
becomes trapped, while skyrmions move readily in the
easy direction.
Recently, Bellizotti Souza \textit{et. al.} \cite{souza_clogging_2022} designed a skyrmion diode device for multiple skyrmions based on an asymmetric
funnel geometry where interacting skyrmions can become clogged.
In the easy-flow direction of the funnel, the skyrmions move
readily and the depinning current is very low; however, the depinning
currents can be much higher for hard direction driving and
reentrant pinning effects can occur in which flowing skyrmions
cease to flow when the drive is increased.
Overall, the skyrmion velocity is much larger for motion along the
easy flow direction than along the hard flow direction,
resulting in a diode effect for multiple skyrmions.

In this work, we use atomistic magnetic simulations to
investigate the dynamics of multiple skyrmions
moving through a periodic array of funnels
constructed using spins with very strong perpendicular magnetic
anisotropy (PMA). The funnel axes are aligned with the $x$ direction and
the transport current is applied along the $-y$ direction. 
We find that for low currents, 
skyrmions can be transported along the funnel with no annihilation.
As the applied current increases, the skyrmions exert pressure on
each other that results in the onset of
skyrmion annihilation.
This annihilation
produces free space inside the sample that
reduces the amount of jamming occurring at the funnel tips
and resulting in an increase in the average skyrmion velocity.
We observe
four distinct dynamic phases as a function of increasing current:
(i) a pinned phase, (ii) motion without annihilation,
(iii) motion with partial annihilation, and (iv) complete annihilation.
The funnel opening plays a major role in the dynamics,
with smaller openings hindering the skyrmion motion due to
a strong bottleneck effect, while larger
openings facilitate the skyrmion motion.
By selecting the right combination of applied external current magnitude
and funnel opening geometry, it is possible to
produce a device that can rapidly transport skyrmions without annihilation.

\section{Simulation}

\begin{figure}[!htb]
   \centering
   \includegraphics[width=\columnwidth]{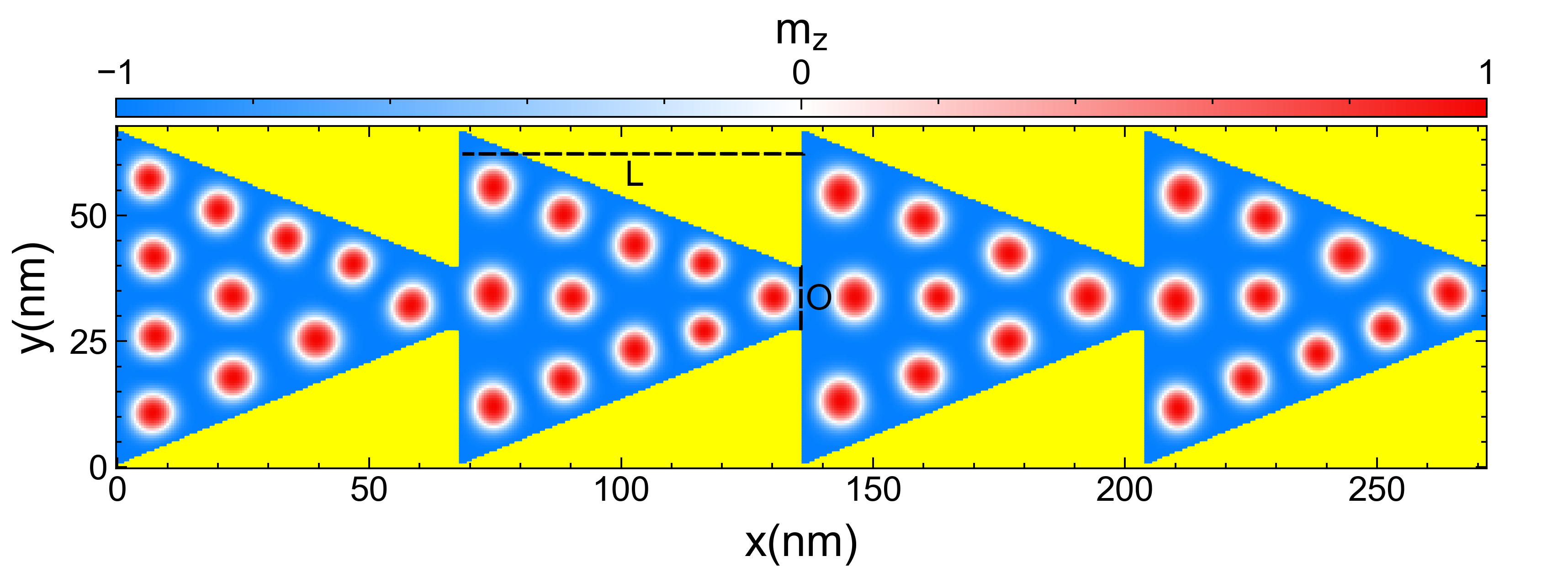}
   \caption{Image of the sample geometry. The funnel array is constructed
     by lining the upper and lower edges of the sample with an asymmetric
     sawtooth arrangement of rigid spins (yellow). 
     The funnel has length $L$ and opening size $O$ as indicated in the figure.
     Inside the funnel,
     color ranging from blue to red indicates the magnitude of the
     out-of-plane magnetization $m_z$. Skyrmions are visible as red circles.}
   \label{fig1}
\end{figure}

We simulate a ferromagnetic thin film that supports Ne{\' e}l skyrmions
at $T=0$K under the influence of a  magnetic field
applied along the $z$ direction, perpendicular to the film plane.
The film has dimensions of
272 nm $\times$ 68 nm and contains rigid spins that form an asymmetric
sawtooth arrangement running parallel to the $x$ axis
along the upper and lower edges of the film, as illustrated
in Fig.~\ref{fig1}. Each funnel plaquette has a length, in the $x$ direction, $L$ and opening $O$.
Throughout the simulation we fix the funnel length as $L = 68$ nm and vary $O$.

We use an atomistic model for the simulations \cite{evans_atomistic_2018} to investigate the skyrmion spin textures in detail.
The following Hamiltonian governs the spin dynamics
\cite{iwasaki_universal_2013,iwasaki_current-induced_2013,seki_skyrmions_2016}:
\begin{equation}\label{Eq1}
\fl\mathscr{H}=-\sum_{i, j\in N}J_{i,j}\mathbf{m}_i\cdot\mathbf{m}_j
               -\sum_{i, j\in N}\mathbf{D}_{i,j}\cdot\left(\mathbf{m}_i\times\mathbf{m}_j\right)
              -\sum_{i}\mu\mathbf{H}\cdot\mathbf{m}_i
              -\sum_{i}K_1\left(\mathbf{m}_i\cdot\hat{\mathbf{z}}\right)^2
\end{equation}

The first term on the right side is the exchange interaction between nearest neighbors,
which compose the set $N$. Here, we consider a square lattice of spins with lattice constant $a$ and exchange constant $J_{i,j}$
between spins $i$ and $j$.
The second term is the Dzyaloshinskii-Moriya interaction,
where $\mathbf{D}_{i,j}$ is
the Dzyaloshinskii-Moriya vector between spins $i$ and $j$ for thin films.
The third term is the Zeeman interaction, with
atomic magnetic moment $\mu$, and the last term
is the sample easy-axis anisotropy where $K_1$ is the anisotropy strength.
We neglect long-range dipolar interactions since they are
expected to be very small for ultrathin films\cite{paul_role_2020}.
The funnel walls are composed of rigid spins
with $\mathbf{m}_F=-\mathbf{\hat{z}}$ fixed over the set $F$
of spins comprising the region outside the funnel.
The rigid spins in magnetic walls are a simplified representation
of spins with extremely high anisotropy. For example, if we replace
the rigid spins by spins that have
$K_1 = 5J$, we achieve the same results shown in this work.

The time evolution of the spins is given by the LLG equation augmented with the adiabatic spin-transfer torque (STT):
\cite{gilbert_phenomenological_2004,slonczewski_current-driven_1996}:
\begin{equation}\label{Eq2}
        \frac{d\mathbf{m}_i}{dt}=
        -\gamma\mathbf{m}_i\times\mathbf{H}^\text{eff}_i
        +\alpha \mathbf{m}_i\times\frac{d\mathbf{m}_i}{dt}
        +\frac{pa^3}{2e}\left(\mathbf{j}\cdot{\nabla}\right)\mathbf{m}_i
\end{equation}

Here $\gamma$ is the gyromagnetic ratio given by $\gamma=g\mu_B/\hbar$,
$\mathbf{H}^\text{eff}_i=-\frac{1}{\mu}\frac{\partial \mathscr{H}}{\partial \mathbf{m}_i}$ is the effective field
including all interactions of the Hamiltonian, $\alpha$ 
is the Gilbert damping parameter, and the last term corresponds to the STT current,
where $p$ is the polarization, $e$ is the electron  charge, $a$ is the lattice constant, and $\mathbf{j}$ the applied current density.
This is called the spin-transfer-torque term, and it includes the assumption
that the conduction electron spins are parallel to the local magnetic moments $\mathbf{m}$ \cite{iwasaki_universal_2013, zang_dynamics_2011}.
We do not include
non-adiabatic spin transfer torque
since it does not appreciably affect the dynamics
of rigid nanoscale skyrmion quasiparticles at small driving forces
\cite{litzius_skyrmion_2017}.
We apply the current along the $-y$ direction, $\mathbf{j}=-j\mathbf{\hat{y}}$.
The direction was chosen based on the interaction of skyrmions with rigid magnetic walls that results in a boosted
skyrmion velocity along the wall
\cite{Juge21,souza_clogging_2022,zhang_structural_2022,zhang_particle-like_2022, bellizotti_souza_spontaneous_2023}.

We measure the number of skyrmions in the sample using the topological charge, where each skyrmion gives a charge of $Q=\pm1$ depending on the 
direction of the applied field. Since we are considering a discrete lattice, the topological charge calculations are done using a lattice-based 
implementation, which is described in detail in Ref.~\cite{kim_quantifying_2020}. Variations in the topological charge calculations due to 
inaccuracies in finite-difference derivatives can be reduced using the
lattice-based approach, especially when nucleation, annihilation, and thermal fluctuations are considered.

The skyrmion velocities are calculated using the emergent electromagnetic field \cite{schulz_emergent_2012,seki_skyrmions_2016}:
\begin{equation}
    E^\text{em}_i=\frac{\hbar}{e}\mathbf{m}\cdot(\partial_i\mathbf{m}\times\partial_t\mathbf{m})
    \;\;,\;\;
    B^\text{em}_i=\frac{\hbar}{2e}\varepsilon_{ijk}\mathbf{m}\cdot(\partial_j\mathbf{m}\times\partial_k\mathbf{m})
\end{equation}
where $\varepsilon_{ijk}$ is the totally anti-symmetric tensor.
The drift velocity $\mathbf{v}_d$ is then computed according to $\mathbf{E}^\text{em}=-\mathbf{v}_d\times\mathbf{B}^\text{em}$ \cite{schulz_emergent_2012,seki_skyrmions_2016}.

Throughout the simulation, we fix $\alpha=0.04$, $p=-1.0$ and $a=0.5$ nm.
We use the following material parameters:
$J=1$ meV, $D=0.18J$, and $K_1=0.01J$.
The applied magnetic field is $\mu\mathbf{H}=-0.5\left(D^2/J\right)\hat{\mathbf{z}}$, enabling the formation of a skyrmion phase \cite{iwasaki_universal_2013,seki_skyrmions_2016}.

We initialize each simulation by placing
randomized spin orientations throughout the sample that
slowly stabilized into the described spin textures after
several relaxation steps
where we use the LLG equation without the spin current term
($\mathbf{j}=\mathbf{0}$).
We relax the system for $3\times10^7$ time steps
in order to guarantee that a steady solution has been reached.
Once the initial state has been prepared,
we introduce a finite spin current and begin to
measure the dynamics of the system.
For the calculations we use a Runge-Kutta fourth-order integration
method was used where time is normalized by $t=(\hbar/J)\tau$
and the spin current
is given by $\mathbf{j}=(2eJ/a^2\hbar)\mathbf{j}^\prime$.
Here, $\tau$ and $\mathbf{j}^\prime$ are the normalized dimensionless units from the code 
that are converted back to the correct dimension units.

\section{Skyrmion Transport}

\begin{figure}[!htb]
   \centering
   \includegraphics[width=0.8\columnwidth]{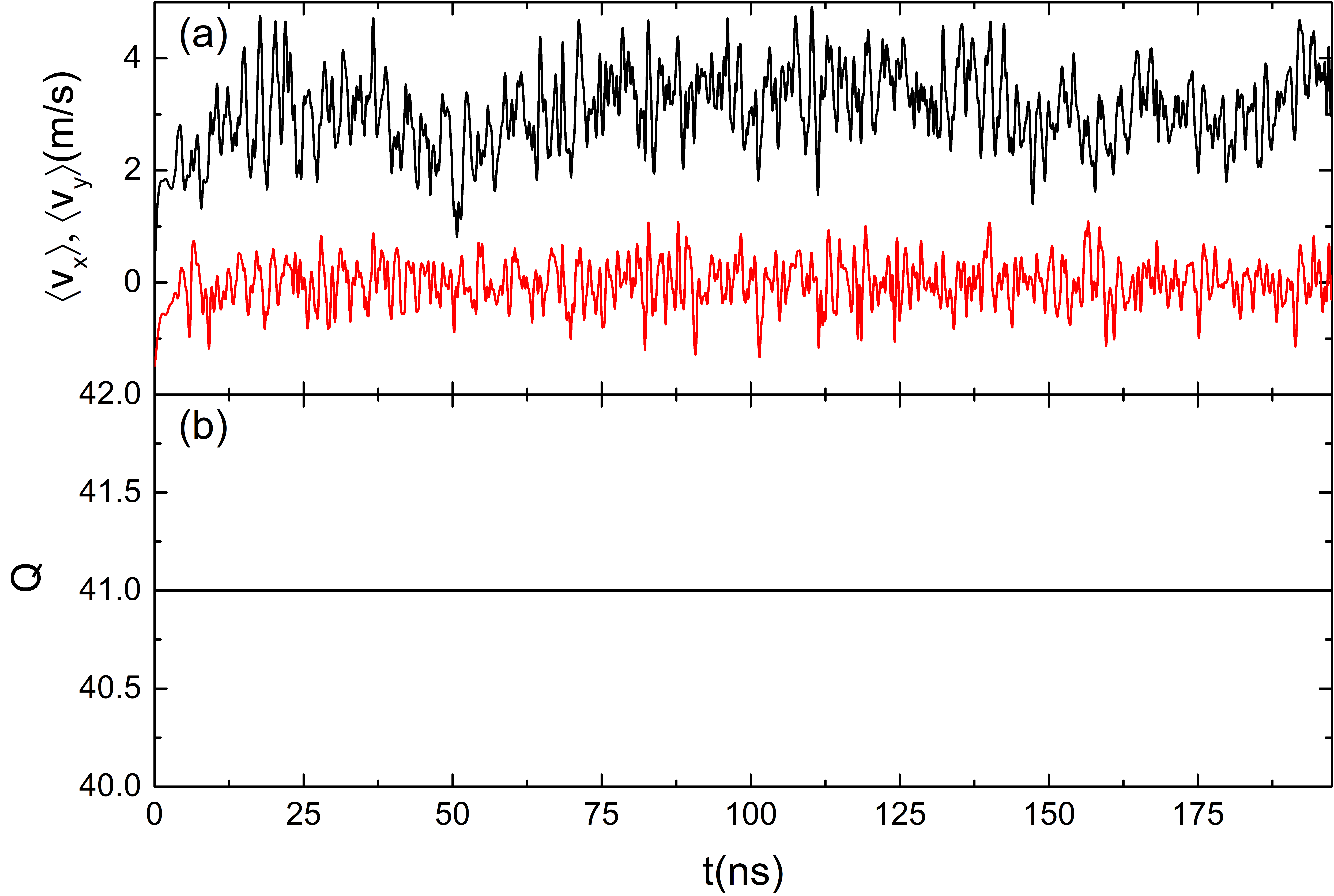}
   \caption{ (a)
     Velocity signals $\left\langle v_x\right\rangle$ (black) and $\left\langle v_y\right\rangle$ (red) vs time $t$ and (b)
   the corresponding topological charge $Q$ vs $t$ for
   a system with $j=0.39\times10^{10}$A/m$^2$ and $O=12.5$nm.
   Both $\left\langle v_x\right\rangle$ and
   $\left\langle v_y\right\rangle$ have noisy signatures.
   The velocity signal oscillates around $\left\langle v_x\right\rangle \approx 3$ m/s and $\left\langle v_y\right\rangle \approx 0$ m/s,
   signifying net motion along the $x$ direction parallel to the
   funnel axis along with oscillations without net motion along
   the $y$ direction.
   The topological charge $Q = 41$ is constant throughout the simulation, indicating that no annihilation is occurring.}
    \label{fig2}
\end{figure}

We first consider a system with
funnel openings of size $O=12.5$ nm
under an applied current of $j=0.39\times10^{10}$A/m$^2$, which
is weak enough that no skyrmion annihilation occurs.
In Fig.~\ref{fig2} we plot the skyrmion average velocities
$\left\langle v_x\right\rangle = |Q|^{-1}\sum_i^{|Q|} {\bf v}_d \cdot {\bf \hat x}$
and
$\left\langle v_y\right\rangle = |Q|^{-1}\sum_i^{|Q|} {\bf v}_d \cdot {\bf \hat y}$
along with the topological charge $Q$
as a function of  simulation time $t$.
In Fig.~\ref{fig2}(a), both velocity signals exhibit a
noisy behavior resulting from the skyrmion-skyrmion
and skyrmion-wall interactions inside the funnel plaquettes.
The skyrmions can deform in size inside the plaquettes, 
creating a turbulent motion with noisy velocity
signatures at different average velocities of
$\left\langle v_x\right\rangle \approx 3$ m/s
and $\left\langle v_y\right\rangle \approx 0$ m/s.
This indicates that the skyrmions undergo
a net motion along the funnel array and
travel from one funnel plaquette to the other along the $x$ direction.
In contrast, there is no net motion in the $y$ direction
but only oscillations in the skyrmion velocity.
In Fig.~\ref{fig2}(b), the topological charge is stabilized at $Q=41$,
indicating that 41 skyrmions are present in the sample and that there is
no annihilation.

\begin{figure}
    \centering
    \includegraphics[width=0.8\columnwidth]{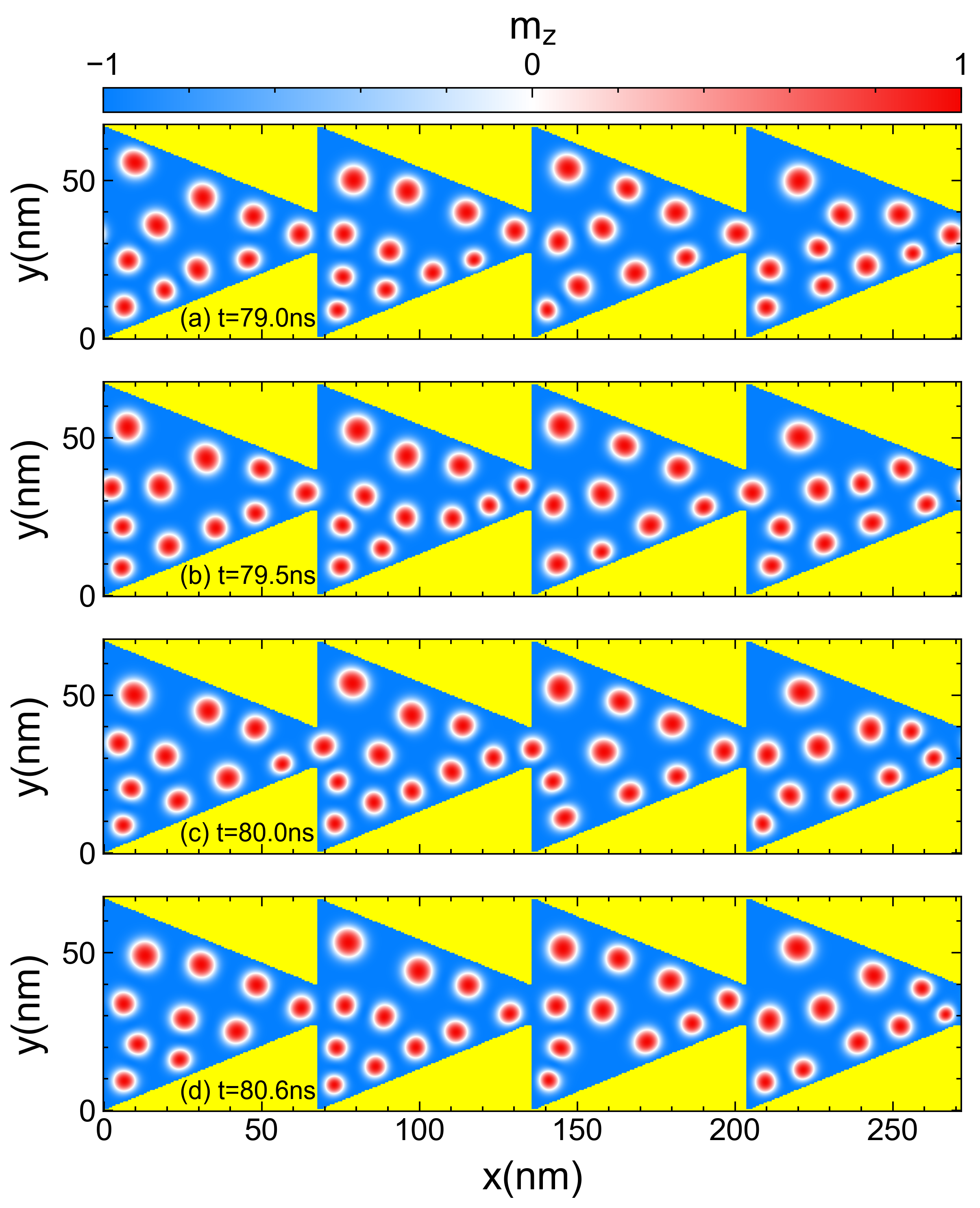}
    \caption{Snapshots of the skyrmion configurations at different
    times illustrating skyrmion motion through the funnel tips
    in a system with $j=0.39\times10^{10}$A/m$^2$ and $O=12.5$ nm.
    (a) $t=79.0$ ns. (b) $t=79.5$ ns. (c) $t=80.0$ ns.
    (d) $t=80.6$ns. 
    }
    \label{fig3}
\end{figure}

In Fig.~\ref{fig3} we show representative snapshots
illustrating the skyrmion motion for the sample from Fig.~\ref{fig2}.
Since the external drive is applied along
the $-y$ direction,
the skyrmions in the upper part of the funnels are larger
and the skyrmions in the lower
part of the funnels are smaller and more crowded.
Compression of the skyrmions against the wall causes them to
slide in the $+x$ direction and slip through the funnel opening
into the next plaquette to the right.
The funnel opening presents a barrier for skyrmions on both the right
and left sides of the opening, but skyrmions approaching from the
left are able to jump over this barrier and move
with high velocity into the neighboring plaquette.
These skyrmions experience
a bottleneck effect, as a result of which only one skyrmion can pass
through the opening at a time.
During the transition from
one funnel plaquette to the other, the skyrmion deforms and shrinks
in order to pass through the
funnel opening.
Interactions with the opening and other skyrmions generate
excitations within the skyrmions,
resulting in the noisy behavior
found in Fig.~\ref{fig2}.
The skyrmions in the upper and lower corners
of the funnel remain mostly immobile,
undergoing deformations and oscillations but no
net motion.
The Supplemental Material includes a video illustrating
the motion inside the funnels.
The overall behavior is very similar to what was observed in
Ref.~\cite{souza_clogging_2022}, where a particle-based model
was used to describe
the jamming and clogging of skyrmions in funnel arrays.
Due to the limitations of the particle model, skyrmion
deformations, excitations, and annihilation were not included in the
previous work.
Here, we show
in greater detail that skyrmion deformations produced by interactions
with the magnetic walls and with other skyrmions
generate internal excitations
that result in the appearance of a a noisy velocity signal.

\section{Skyrmion Annihilation}

\begin{figure}[!htb]
  \centering
  \includegraphics[width=0.8\columnwidth]{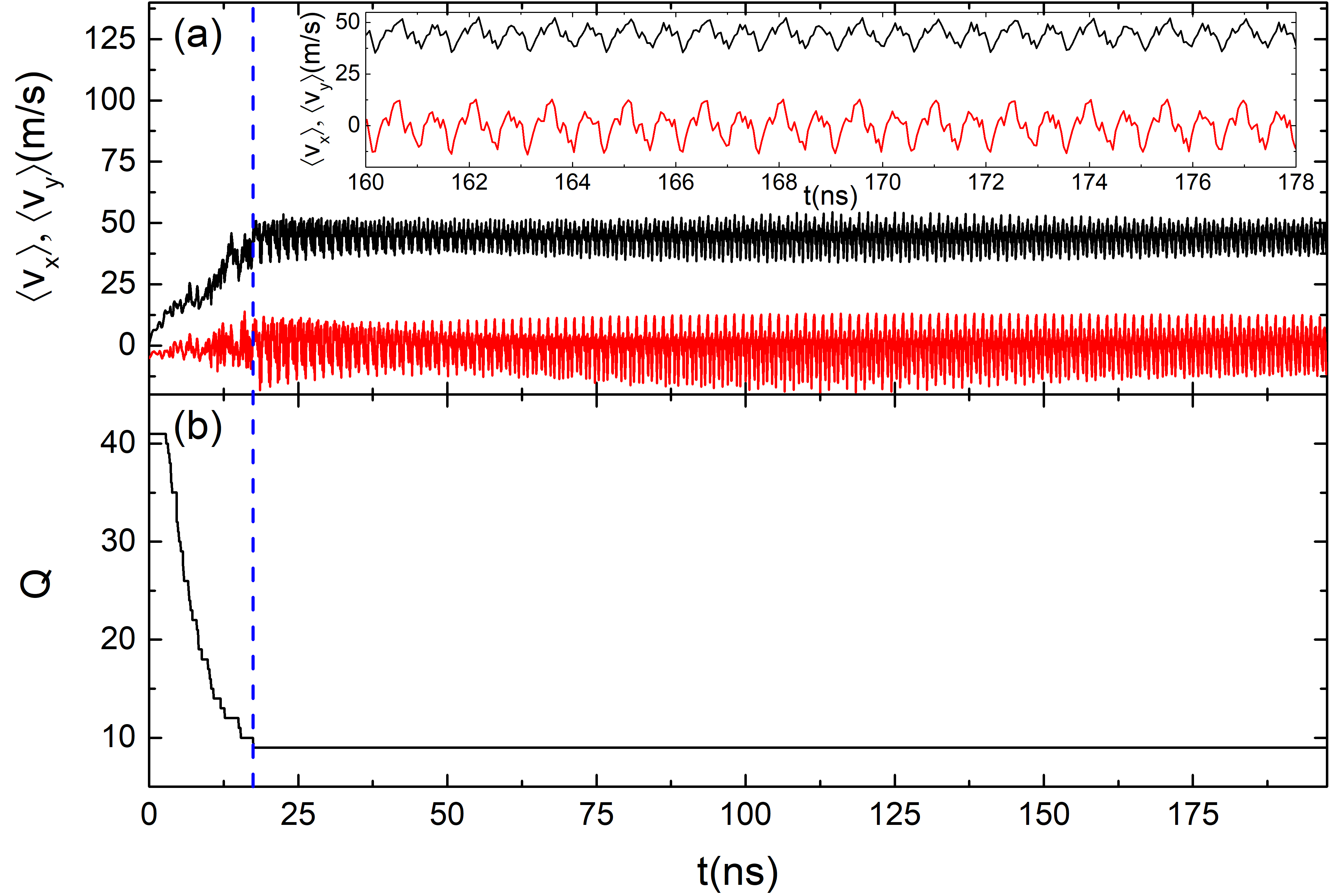}
  \caption{(a) Velocity signals
  $\left\langle v_x\right\rangle$ (black) and
  $\left\langle v_y\right\rangle$ (red)
  vs $t$ and (b) the corresponding topological
  charge $Q$ vs $t$ for a system with
  $j=1.36\times10^{10}$ A/m$^2$ and $O=12.5$ nm. The
  blue dashed line at $t=17$ ns indicates
  the end of the annihilation process.
  The inset in panel (a) shows a detail of the velocity time
  series from the main panel.
  }
    \label{fig4}
\end{figure}

We next describe the influence of skyrmion annihilation
on the skyrmion transport.
When the applied current is low, skyrmions can
move slowly along the funnel arrays
and undergo deformation and jamming but no annihilation.
In Fig.~\ref{fig4}(a) we plot$\left\langle v_x\right\rangle$
and $\left\langle v_y\right\rangle$ versus $t$ for a system with
$O=12.5$ nm at a higher current of
$j=1.36\times10^{10}$ A/m$^2$, while in Fig.~\ref{fig4}(b)
we show the corresponding topological charge $Q$
versus $t$.
For $t=0$, there are 41 skyrmions present and $Q=41$,
just as in Section 3 where no annihilation occurred.
Here the current density is large enough that skyrmions begin to
annihilate until a final steady value of
$Q=Q_f=9$ is reached.
The blue dashed line at $t=17$ ns marks the moment at
which the annihilation process ends, indicated by the fact that the number
of skyrmions in the sample remains constant.
During the annihilation period,
the velocity signal $\left\langle v_x\right\rangle$ is generally
increasing with time.
This velocity increase results because the destruction of skyrmions
increases the amount of free space inside each plaquette,
decreasing the effectiveness of jamming and enabling greater skyrmion
mobility along the $x$ direction.
There are two main drivers for skyrmion annihilation in our system.
(i) Pressure from the applied current pushes the skyrmions
toward the lower funnel wall.
(ii) Skyrmions that are pushed by the drive also exert pressure
on the skyrmions that are closest to the wall.
The annihilation process is very similar to what has been
observed recently for skyrmions interacting with magnetic walls
\cite{bellizotti_souza_spontaneous_2023}, and consists of
heavy deformation of the skyrmion followed by shrinking to annihilation.
Since fewer skyrmions are present in the sample after the annihilation
is complete, and these skyrmions have shrunk in size
due to the pressure from the applied current, the remaining skyrmions
are able to pass more easily through the funnel opening and the net
velocity along the easy flow direction is increased.
In addition, the surviving skyrmions interact
strongly with the wall and experience a Magnus velocity
boost \cite{Juge21,bellizotti_souza_magnus_2022}.

\begin{figure}[!htb]
  \centering
  \includegraphics[width=0.8\columnwidth]{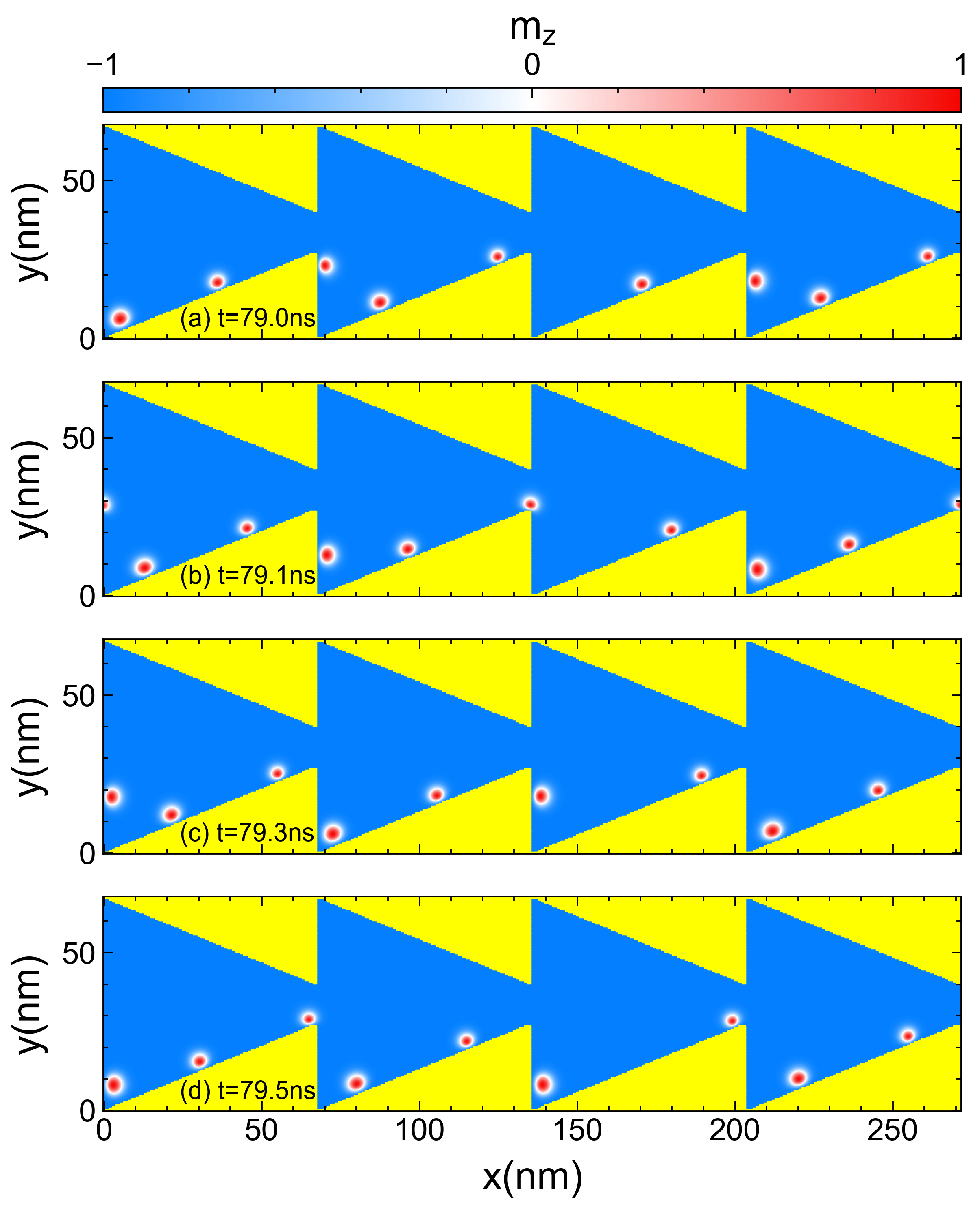}
  \caption{Snapshots of the skyrmion configurations at
  different times illustrating periodic skyrmion motion
  in the system containing $Q_f=9$ skyrmions
  with $j=1.36\times10^{10}$ A/m$^2$ and 
  $O=12.5$ nm for times after the end
  of the annihilation process.
  (a) $t=79.0$ ns. (b) $t=79.1$ ns. (c) $t=79.3$ ns.
  (d) $t=79.5$ns. 
    }
    \label{fig5}
\end{figure}

After the annihilation process has ended, the velocity
no longer increases but instead exhibits a periodic
signal, as highlighted in the inset of Fig \ref{fig4}(a)
for the interval $160\leq t\leq 178$ ns.
The velocity peaks at moments when
most of the skyrmions are interacting with the lower walls of
the funnel, where the Magnus boost effect can occur.
Velocity minima occur each time a surviving skyrmion reaches
the funnel tip and must overcome the barrier to reach the next
plaquette.
In this regime, the number of surviving skyrmions is sufficiently small
that the skyrmion-skyrmion distance is large, rendering the interactions
between skyrmions negligible and causing the behavior to be dominated
by interactions with the walls, unlike what we observed in
Section 3.
The velocities oscillate around average values of
$\left\langle v_x\right\rangle \approx 45$ m/s
and $\left\langle v_y\right\rangle \approx 0$ m/s.
The oscillation frequency
depends on both the funnel size, $L$ and the current density, $j$,
with the frequency increasing for shorter funnels and decreasing
for lower currents.
Periodic behavior only occurs for systems
in which each funnel plaquette hosts
a steady state number of skyrmions that is small enough to allow
all of the skyrmions to flow along the funnel wall,
permitting interactions with the current and the wall
to dominate over skyrmion-skyrmion interactions.
Snapshots of the time evolution of skyrmions
undergoing periodic motion
appear in Fig.~\ref{fig5}, while the
Supplemental Material
includes a video illustrating the annihilation process for
$t<17$ ns and the periodic skyrmion motion for $t>17$ ns.

\section{Transport Current}

In the previous sections, we demonstrated that the applied current
density can significantly affect the skyrmion behavior and determines
whether annihilation processes occur.
We next investigate in more detail
how the magnitude of the current density can modify the skyrmion dynamics.
We fix the funnel opening to $O=12.5$ nm and vary the applied current
over the interval
$0.03\times10^{10}$ A/m$^2\leq j\leq 1.58\times10^{10}$A/m$^2$.

\begin{figure}[!htb]
  \centering
  \includegraphics[width=0.8\columnwidth]{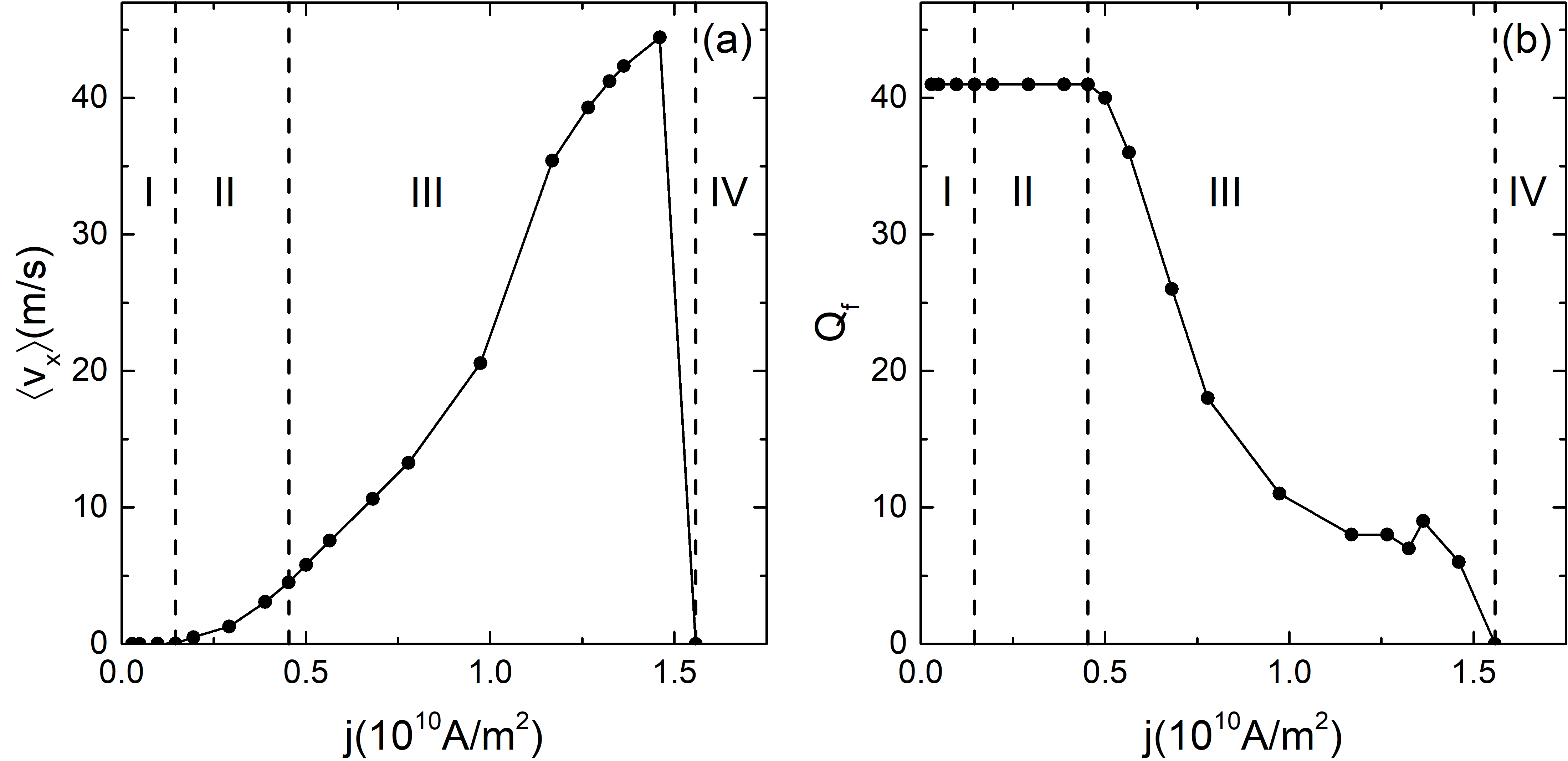}
  \caption{(a)$\left\langle v_x\right\rangle$ and
  (b) final topological charge $Q_f$ versus
 applied current $j$ in a system with fixed
 funnel opening size $O=12.5$nm.
 The dashed black lines indicate the boundaries of the
 four dynamic phases:
 I, no motion; II, motion without annihilation;
 III, motion with annihilation, and IV, complete annihilation
 of all skyrmions.
 There is
 a monotonic increase in $\langle v_x\rangle$ with $j$
 until all of the skyrmions in the
 sample are annihilated, which occurs for
 $j\geq 1.56\times10^{10}$ A/m$^2$.
 $Q_f$ exhibits a stable value in regions I and II,
 followed by a monotonic decrease with increasing $j$ until $Q_f=0$.}
    \label{fig6}
\end{figure}

In Fig.~\ref{fig6}(a) we plot $\left\langle v_x\right\rangle$,
the velocity along the funnel axis,
versus the applied current density $j$.
We find a monotonic increase in $\langle v_x\rangle$
with increasing applied current up to
$j= 1.45\times10^{10}$ A/m$^2$.
In the pinned region, labeled I,
the skyrmions exhibit no net motion and are smoothly compressed
against the wall. Region II, where motion without annihilation occurs,
begins when the skyrmions depinning for $j=0.15\times10^{10}$ A/m$^2$.
The plot of the final topological charge $Q_f$ versus $j$ in
Fig.~\ref{fig6}(b) shows that in regions I and II,
the topological charge remains constant at $Q_f = 41$.
We note that at each value of $j$,
the skyrmions are first initiated and stabilized in a
sample with $j=0$, and then the external drive is applied, which may
modify the topological charge as shown
in Fig.~\ref{fig4}(b). Thus we report only the final stable topological
charge value $Q_f$ in Fig.~\ref{fig6}(b).
In region III, for $j > 0.45\times10^{10}$ A/m$^2$, the skyrmions are
moving but an annihilation process occurs, and the number of
skyrmions in the sample decreases with increasing $j$, resulting in a
corresponding increase in the average skyrmion velocity.
There are some small oscillations in the topological charge
$Q_f$ near $j \approx 1.25\times10^{10}$A/m$^2$, but this is a result of
the statistics of the
small number of skyrmions present in the sample for these large
currents.
Increasing the current further causes the skyrmions to
compress more strongly against
bottom wall, shrinking the skyrmions
and initiating the annihilation process described in detail
in the previous section.
The monotonic increase
of $\left\langle v_x\right\rangle$
with $j$ in regions II and III is a consequence of the Magnus velocity
boost effect caused by skyrmions sliding along the wall.
In region IV,
for $j > 1.45\times10^{10}$ A/m$^2$, there is complete annihilation of all
skyrmions in the sample and
the skyrmion average velocity decreases sharply to zero.

\section{Funnel Opening}

Up until this point we have considered
skyrmions interacting with a funnel array of fixed
opening size $O=12.5$ nm.
We next show how changing the opening size
can significantly influence the dynamics
since it controls the bottleneck effect.
Smaller openings increase the bottleneck effect while
larger openings reduce it, and thus, an understanding
of how the funnel opening influences the skyrmion transport
and annihilation processes through the bottleneck
effect is essential to better understand the skyrmion dynamics.
We vary the funnel opening over the range
5 nm$\leq O\leq 25$ nm for applied currents
ranging from
$0.03\times10^{10}$ A/m$^2\leq j\leq 1.58\times10^{10}$ A/m$^2$.

\begin{figure}[!htb]
    \centering
    \includegraphics[width=0.8\columnwidth]{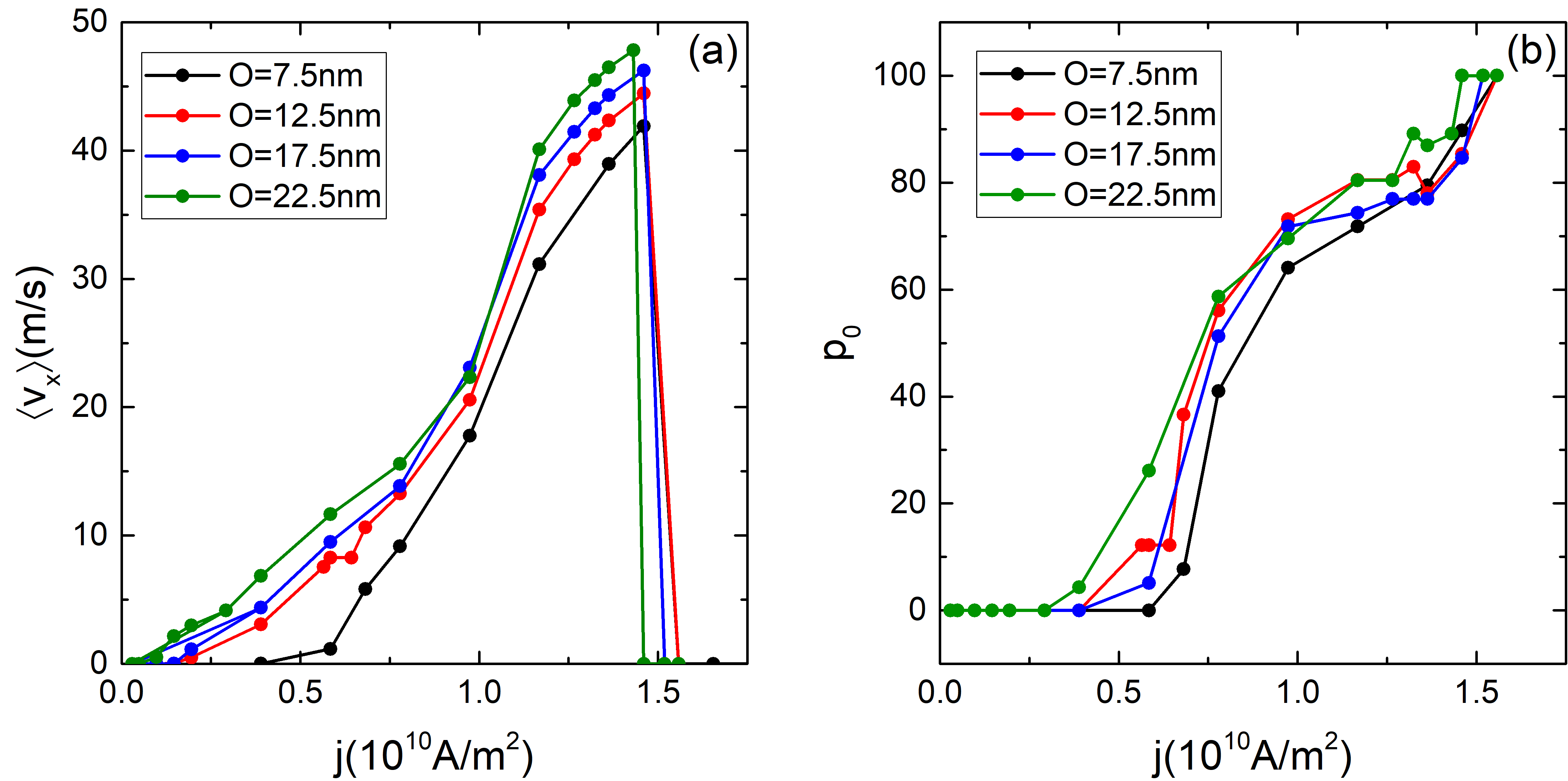}
    \caption{(a) Skyrmion average velocity
    $\left\langle v_x\right\rangle$ and (b)
    the extinction probability $p_0$ giving the
    percentage of skyrmions annihilated during the simulation
    vs applied current $j$ for four different funnel opening
    sizes, $O$: $O=7.5$ nm (black),
    $O=12.5$ nm (red), $O=17.5$ nm (blue),
    and $O=22.5$ nm (green).
    }
    \label{fig7}
\end{figure}

\begin{figure}[!htb]
    \centering
    \includegraphics[width=0.8\columnwidth]{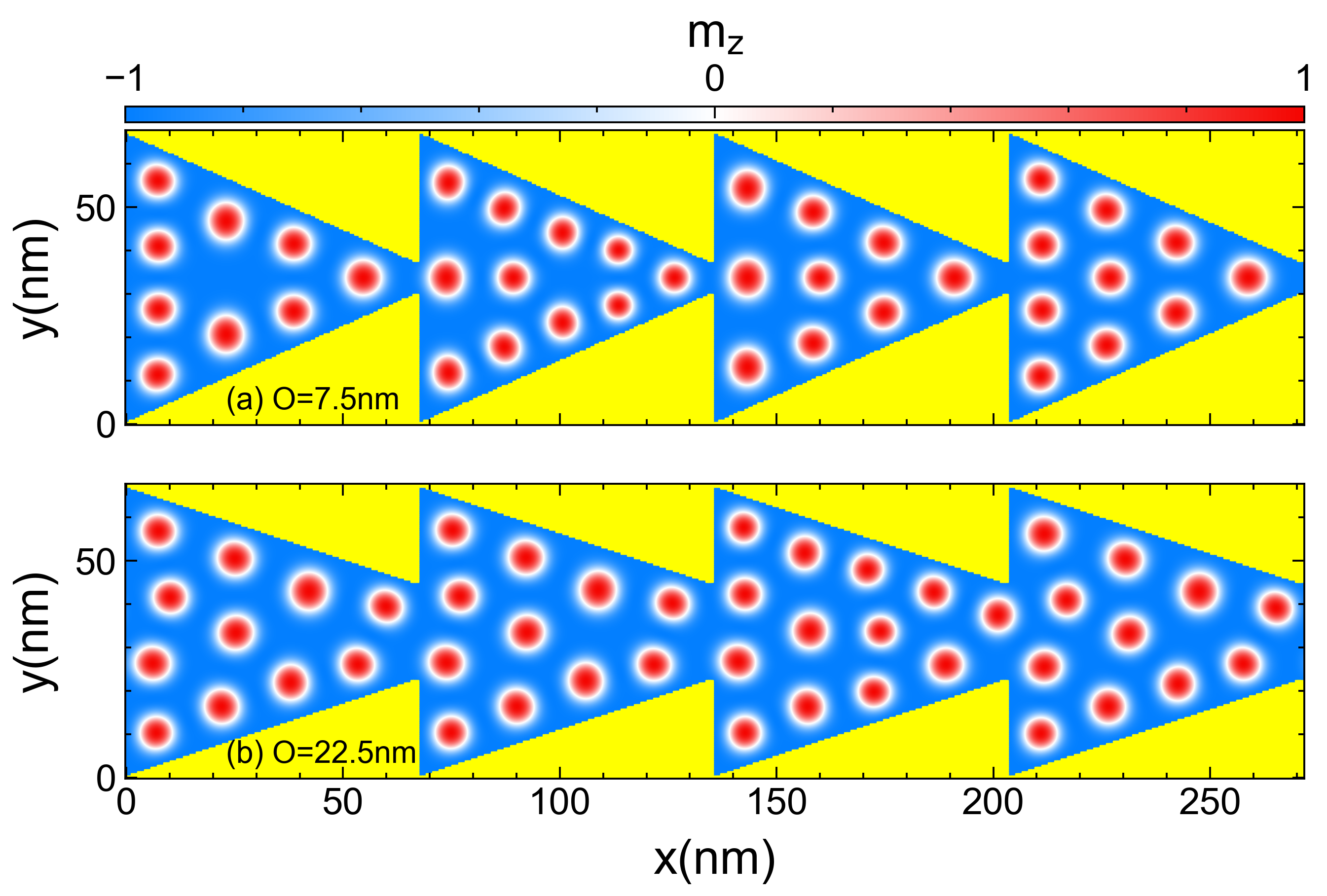}
    \caption{Snapshots of the skyrmion ground state configurations
      obtained after initializing the system with $j=0$ for different
      funnel opening sizes of (a) $O=7.5$ nm and (b) $O=22.25$ nm.
      In (a) there are $Q=39$ skyrmions and in (b)
      there are $Q=46$ skyrmions. Fewer skyrmions can be stabilized
      when the funnel opening size is reduced, which affects the
      dynamics once a current is applied.
    }
    \label{fig8}
\end{figure}

In Fig.~\ref{fig7} we plot the average velocity,
$\left\langle v_x\right\rangle$, and the annihilation probability, $p_0$
versus applied current density $j$ for different funnel opening sizes.
Here $p_0$ describes the fraction of skyrmions originally present in the
sample that are annihilated by the time steady state is reached.
The general behavior of both $\langle v_x\rangle$ and $p_0$ is not
changed when the funnel opening size is varied; instead,
the values of $j$ at which the different regions occur shifts.
as $O$ changes.
The depinning threshold drops to lower values of $j$ as $O$ increases.
For example, for $O=7.5$ nm, we find
a depinning threshold of $j=0.39\times10^{10}$ A/m$^2$ that
is much larger than the depinning threshold $j=0.05\times10^{10}$ A/m$^2$
of the $O=22.5$ nm sample.
Smaller funnel openings enhance the bottleneck
effect, making it necessary for more energy to be expended in order to
deform the skyrmions enough to permit them to pass through the energy
barrier separating adjacent plaquettes.
As the funnel openings become larger,
less energy is necessary to depin the skyrmions, leading to
reduced depinning thresholds.
Figure \ref{fig7}(b) shows the extinction probability $p_0$ or the
fraction of skyrmions annihilated during the simulation
versus $j$.
Minor differences appear as the funnel opening changes,
but the general behavior remains the same.
There is an annihilation-free plateau at low currents spanning
regions I and II where the skyrmions either remain pinned or
can flow while being conserved, followed by region III behavior of
increasing annihilation
up to complete skyrmion annihilation in region IV at high drives.
The onset of the annihilation process shifts
to lower values of $j$ with increasing funnel opening size $O$.
Larger funnel openings permit
more skyrmions to remain stabilized in the sample since there is
an expanded space for skyrmion motion
and the applied magnetic field is held at a fixed value.
In Fig.~\ref{fig8} we illustrate the skyrmion ground states
obtained after initialization at $j=0$
for systems with $O=7.5$ nm and $O=22.5$ nm.
For $O=22.5$ nm in Fig.~\ref{fig8}(a), the wider
funnel tips provide
an increased space for accommodating skyrmions in the sample,
resulting in the stabilization of $Q=46$ skyrmions. On the other hand, for
$O=7.5$ nm, less space is available and only
$Q=39$ skyrmions are stabilized.
As a consequence, when an external drive is applied to the
$O=22.5$ nm sample from
Fig.~\ref{fig8}(b),
there is a greater compression of the skyrmions against the bottom wall
of the funnel.
This increased pressure favors enhanced skyrmion annihilation and a
larger value of $p_0$ when the funnel opening is large.
In contrast, for the smaller $O=7.5$ nm funnel
opening size in Fig.~\ref{fig8}(a),
the decrease in the number of stabilized skyrmions causes a corresponding
decrease in 
the pressure exerted on the skyrmions adjacent to the lower funnel wall
when a finite driving force is applied.
Finally, when the funnel openings are larger, the orientation of the long
funnel walls comes closer to being parallel to the $x$ axis and perpendicular
to the drive that is applied along the $-y$ direction.
This makes the lower funnel wall more effective at exerting pressure
against the bottom row of driven skyrmions and lowers the threshold
current at which skyrmion annihilation begins.

\begin{figure}[!htb]
    \centering
    \includegraphics[width=0.4\columnwidth]{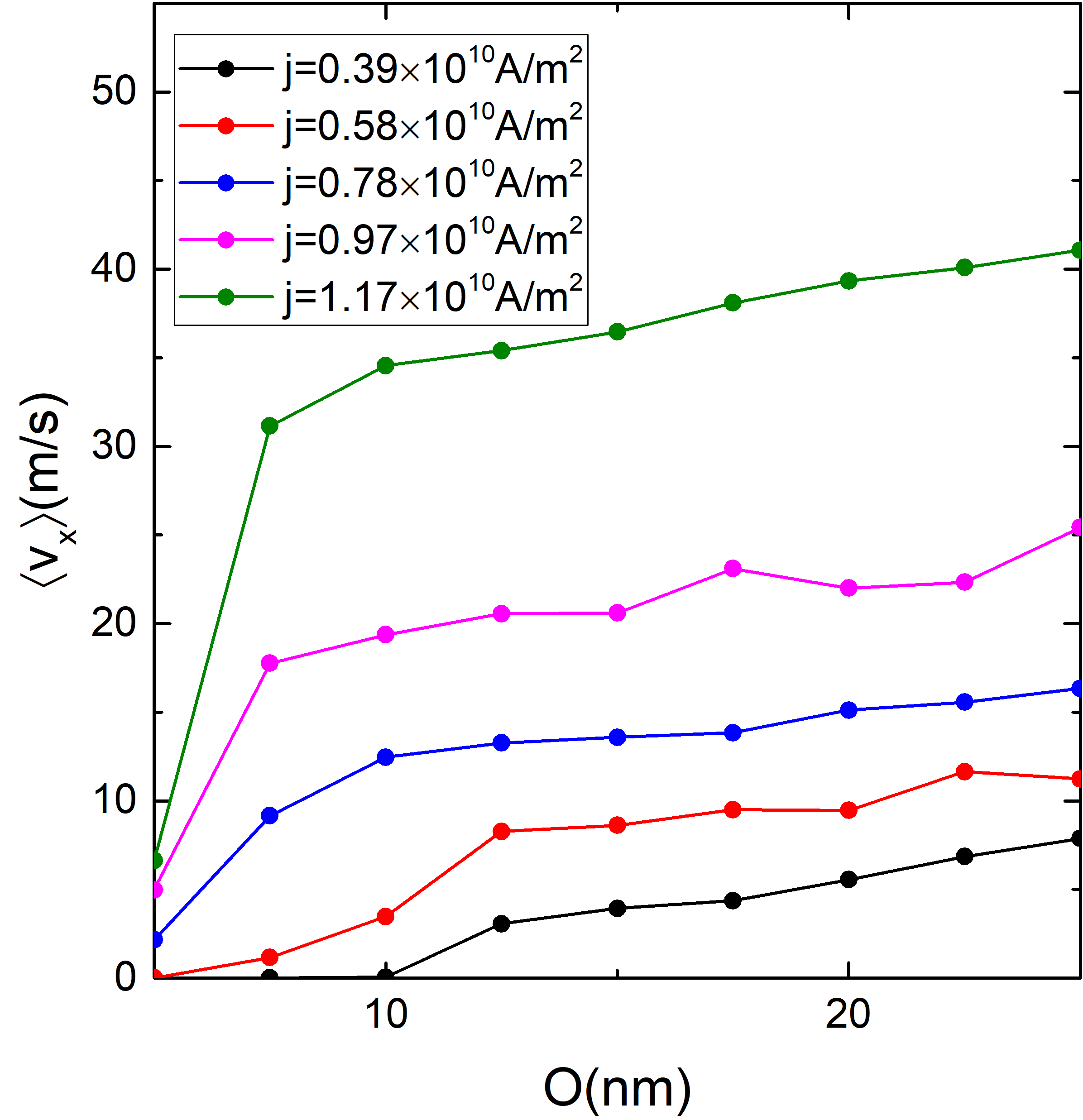}
    \caption{$\left\langle v_x\right\rangle$
    vs funnel opening size $O$ for different applied currents of
    $j=0.39\time10^{10}$ A/m$^2$ (black),
    $j=0.58\times10^{10}$ A/m$^2$ (red),
    $j=0.78\times10^{10}$ A/m$^2$ (blue),
    $j=0.97\times10^{10}$ A/m$^2$ (magenta), and
    $j=1.17\times10^{10}$ A/m$^2$ (green).}
    \label{fig9}
\end{figure}

In Fig.~\ref{fig9} we plot $\langle v_x\rangle$ versus
the funnel opening size $O$ for
selected values of the applied current in order to illustrate
more clearly
the skyrmion transport velocity is affected by
the funnel opening size.
We find that $\langle v_x\rangle$ monotonically increases with
increasing $O$ 
for all values of the external drive $j$.
The increase in $\langle v_x\rangle$ is the most pronounced
for small values of $O$, while $\langle v_x\rangle$ tends toward
a saturation value as $O$ becomes large.
This saturation value varies according
to the magnitude of the current density, with larger currents
giving larger saturation velocities.
As the funnel openings become very large, 
the funnel structure itself begins to vanish,
leading to the well studied case of
skyrmions in a
nanotrack \cite{purnama_guided_2015,Juge21,sampaio_nucleation_2013}.
For low applied currents, such as $j=0.39\times10^{10}$ A/m$^2$, the
skyrmions cannot be transported through the funnel opening
when $O\leq 10$ nm due to a strong jamming effect.
As the applied current increases, this depinning
threshold shifts to smaller values of $O$.
For example, for $j=0.58\times10^{10}$ A/m$^2$, 
the threshold depinning opening is around $O=5$ nm.

\begin{figure}[!htb]
  \centering
  \includegraphics[width=0.4\columnwidth]{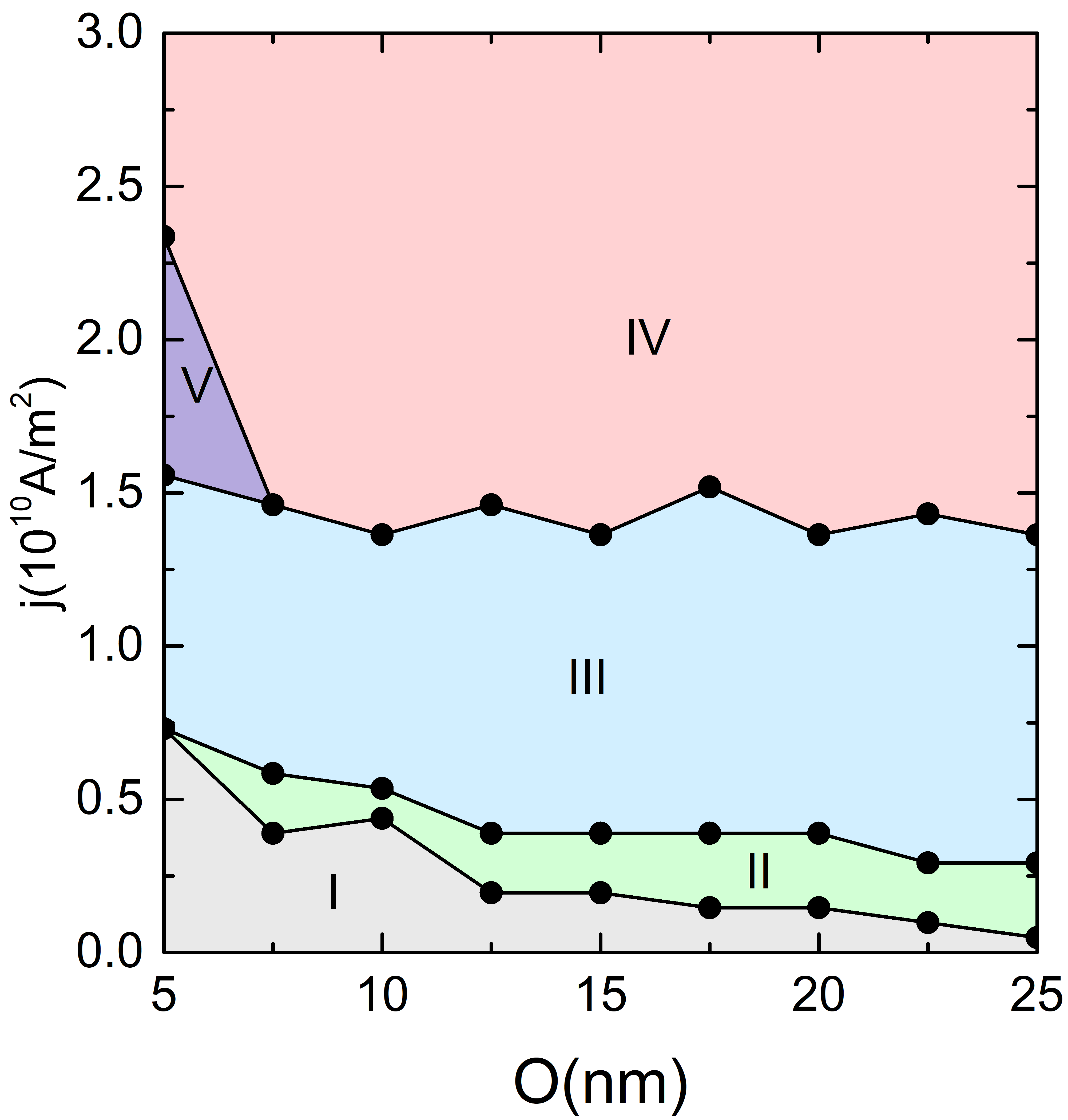}
  \caption{Dynamic phase diagram as a function of applied current
  $j$ versus funnel opening size $O$ highlighting
  five different regions.
  Region I (gray): no motion.
  Region II (green): motion without annihilation.
  Region III (light blue): motion with annihilation.
  Region IV (red): complete annihilation of all skyrmions.
  Region V (purple) is a reentrant pinned phase with no motion.}
    \label{fig10}
\end{figure}

To summarize the general skyrmion transport behavior
in funnel geometries, in Fig.~\ref{fig10} we plot
a dynamic phase diagram
as a function of applied current $j$ versus funnel opening size $O$,
where we highlight five regimes.
Region I is the static or pinned
phase where the skyrmions exhibit no net motion.
It is present for all values of $O$ considered in this work,
but the depinning threshold generally decreases with increasing $O$.
For small $O$, region I becomes more prominent due to
the jamming of skyrmions produced by a
strong bottleneck effect.
In region II, skyrmions are transported along the funnels without annihilation,
and the skyrmion number is conserved.
This region appears only for a narrow window of current and disappears when
$O\leq5$ nm.
When $O=5$ nm, the
funnel opening is too small to permit stable skyrmions to pass, and
a skyrmion that tries to move from one funnel plaquette to another
experiences such a strong compressive deformation that it annihilates
rather than successfully translating to the next plaquette.
Region III consists of skyrmion transport with partial annihilation.
Here, the skyrmions can be transported through the funnels,
but a fraction $p_0>0$ of the skyrmions originally present
are annihilated.
This annihilation results primarily when the applied current produces
pressure on the skyrmions along the bottom wall of the funnel, both
directly and as a result of compression from one or more layers of
skyrmions that are not directly in contact with the bottom wall.
This pressure causes the lower skyrmions to shrink, and eventually these
skyrmions become so strongly deformed that they annihilate.
In Region IV, we find complete annihilation of all skyrmions in the
sample during transport.
The threshold value of $j$ at which this annihilation occurs is
nearly independent of $O$ and falls at 
$j\approx1.5\times10^{10}$ A/m$^2$.
This is expected because the critical current value
that annihilates all skyrmions
depends almost exclusively on the pressure exerted by the external drive
and not on the angle of the long funnel walls or the
strength of the skyrmion-skyrmion
interactions.
For very small funnel openings of $O<7.5$ nm, we find
a reentrant pinned state marked region V.
In this regime, skyrmions begin to annihilate when a drive is applied,
but when only a few skyrmions remain in the sample, they become unable
to pass through the small opening separating neighboring plaquettes,
resulting in a stable pinned configuration.
For thees small funnel openings, the skyrmion transport from one
funnel to the next is primarily driven by collective skyrmion
interactions,
where each skyrmion pushes its neighboring skyrmion through the funnel
tip in a chain reaction process.
When the current reaches the lower boundary of region V,
most of the skyrmions in the sample are annihilated
and this collective motion is destroyed, trapping the skyrmions in
individual funnel plaquettes.

\section{Summary}
Using atomistic simulations, we investigated
the dynamical behavior of skyrmions interacting with a funnel geometry.
under different currents applied along the $-\hat{\mathbf{y}}$ direction,
perpendicular to the funnel axis.
We find that skyrmions can be transported through the funnel
openings due to a combination of
Magnus force and collective effects.
For driving currents just above the depinning threshold,
the skyrmions can be transported without annihilation.
For intermediate drive, the skyrmions continue to move but some skyrmions
are destroyed in an annihilation process that results
from skyrmion-wall and skyrmion-skyrmion interactions.
The external current causes the skyrmions in the upper part of the funnel
to exert pressure on the skyrmions along the bottom of the funnel,
which shrink to the point that they eventually annihilate.
As more skyrmions are annihilated, the collective interactions are
reduced and the remaining skyrmions
move more freely through the funnels.
Under large drives of
$j \approx 1.5\times10^{10}$ A/m$^2$,
the pressure from the driving current is too strong
and all of the skyrmions in the sample are annihilated.

We demonstrate that the size of the funnel opening plays a major role in the
dynamics. It takes more energy to force skyrmions through small funnel
openings since
the skyrmions must deform significantly
in order to squeeze through the opening, and as a result the
onset of skyrmion motion shifts to higher currents as the funnel openings
decrease in size.
For larger funnel openings,
the skyrmions deform less and require less energy to pass through the
opening.
The onset of the annihilation process
also shifts to lower current values as
the funnel opening size increases.
Larger funnel openings provide
a greater amount of space
in which more skyrmions can be stabilized at zero current
since we hold the applied
magnetic field fixed.
When a current is then applied,
the increased skyrmion density in funnels with large openings
increases the current-induced
pressure exerted by skyrmions in the upper part of the sample
on skyrmions along the lower funnel wall.
As a result, the annihilation threshold drops to lower applied currents
in systems with larger funnel openings.

Our findings provide new insights into the role of
collective effects, dynamics, and annihilation processes
for skyrmions in nanodevices
that could be used in future technological applications.
Our results will be of use in developing
new ways to transport skyrmions,
permitting the skyrmions to carry information in a controlled manner.

\section*{Acknowledgements}

This work was supported by the US Department of Energy through the Los Alamos National Laboratory. Los
Alamos National Laboratory is operated by Triad National Security, LLC, for the National Nuclear Security
Administration of the U. S. Department of Energy (Contract No. 892333218NCA000001). 
J.C.B.S and N.P.V acknowledge funding from Fundação de Amparo à Pesquisa do Estado
de São Paulo - FAPESP (Grants 2022/14053-8 and 2017/20976-3 respectively).
We would like to thank Dr. Felipe F. Fanchini for providing the computational resources used in this work. 
These resources were funded by the Fundação de Amparo à Pesquisa do Estado de São Paulo - FAPESP (Grant: 2021/04655-8).

\section*{References}
\bibliographystyle{iopart-num}
\bibliography{refs}

\providecommand{\newblock}{}
\begin{thebibliography}{10}
\expandafter\ifx\csname url\endcsname\relax
  \def\url#1{{\tt #1}}\fi
\expandafter\ifx\csname urlprefix\endcsname\relax\def\urlprefix{URL }\fi
\providecommand{\eprint}[2][]{\url{#2}}

\bibitem{nagaosa_topological_2013}
Nagaosa N and Tokura Y 2013 {\em Nature Nanotechnol.\/} {\bf 8} 899--911

\bibitem{everschor-sitte_perspective_2018}
Everschor-Sitte K, Masell J, Reeve R~M and Kläui M 2018 {\em Journal of Applied Physics\/} {\bf 124} 240901

\bibitem{fert_magnetic_2017}
Fert A, Reyren N and Cros V 2017 {\em Nature Rev. Mater.\/} {\bf 2} 1--15

\bibitem{muhlbauer_skyrmion_2009}
M{\" u}hlbauer S, Binz B, Jonietz F, Pfleiderer C, Rosch A, Neubauer A, Georgii R and B{\" o}ni P 2009 {\em Science\/} {\bf 323} 915--919

\bibitem{yu_real-space_2010}
Yu X~Z, Onose Y, Kanazawa N, Park J~H, Han J~H, Matsui Y, Nagaosa N and Tokura Y 2010 {\em Nature\/} {\bf 465} 901--904

\bibitem{munzer_skyrmion_2010}
Münzer W, Neubauer A, Adams T, Mühlbauer S, Franz C, Jonietz F, Georgii R, Böni P, Pedersen B, Schmidt M, Rosch A and Pfleiderer C 2010 {\em Physical Review B\/} {\bf 81} 041203

\bibitem{pfleiderer_skyrmion_2010}
Pfleiderer C, Adams T, Bauer A, Biberacher W, Binz B, Birkelbach F, Böni P, Franz C, Georgii R, Janoschek M, Jonietz F, Keller T, Ritz R, Mühlbauer S, Münzer W, Neubauer A, Pedersen B and Rosch A 2010 {\em Journal of Physics: Condensed Matter\/} {\bf 22} 164207

\bibitem{yu_near_2011}
Yu X~Z, Kanazawa N, Onose Y, Kimoto K, Zhang W~Z, Ishiwata S, Matsui Y and Tokura Y 2011 {\em Nature Materials\/} {\bf 10} 106--109

\bibitem{dzyaloshinsky_thermodynamic_1958}
Dzyaloshinsky I 1958 {\em Journal of Physics and Chemistry of Solids\/} {\bf 4} 241--255

\bibitem{moriya_anisotropic_1960}
Moriya T 1960 {\em Physical Review\/} {\bf 120} 91--98

\bibitem{schulz_emergent_2012}
Schulz T, Ritz R, Bauer A, Halder M, Wagner M, Franz C, Pfleiderer C, Everschor K, Garst M and Rosch A 2012 {\em Nature Phys.\/} {\bf 8} 301--304

\bibitem{Reichhardt22a}
Reichhardt C, Reichhardt C~J~O and Milosevic M 2022 {\em Rev. Mod. Phys.\/} {\bf 94} 035005

\bibitem{jiang_direct_2017}
Jiang W, Zhang X, Yu G, Zhang W, Wang X, Jungfleisch M~B, Pearson J~E, Cheng X, Heinonen O, Wang K~L, Zhou Y, Hoffmann A and te~Velthuis S~G~E 2017 {\em Nature Phys.\/} {\bf 13} 162--169

\bibitem{Reichhardt15}
Reichhardt C, Ray D and Reichhardt C~J~O 2015 {\em Phys. Rev. Lett.\/} {\bf 114}(21) 217202

\bibitem{litzius_skyrmion_2017}
Litzius K, Lemesh I, Kr{\" u}ger B, Bassirian P, Caretta L, Richter K, B{\" u}ttner F, Sato K, Tretiakov O~A, F{\" o}rster J, Reeve R~M, Weigand M, Bykova I, Stoll H, Sch{\" u}tz G, Beach G~S~D and Kl{\" a}ui M 2017 {\em Nature Phys.\/} {\bf 13} 170--175

\bibitem{zeissler_diameter-independent_2020}
Zeissler K, Finizio S, Barton C, Huxtable A~J, Massey J, Raabe J, Sadovnikov A~V, Nikitov S~A, Brearton R, Hesjedal T, van~der Laan G, Rosamond M~C, Linfield E~H, Burnell G and Marrows C~H 2020 {\em Nature Commun.\/} {\bf 11} 428

\bibitem{zhang_magnetic_2017}
Zhang X, Xia J, Zhao G~P, Liu X and Zhou Y 2017 {\em {IEEE} Transactions on Magnetics\/} {\bf 53} 1--6

\bibitem{reichhardt_quantized_2015}
Reichhardt C, Ray D and Reichhardt C~J~O 2015 {\em Physical Review B\/} {\bf 91} 104426

\bibitem{Ma16}
Ma F, Reichhardt C, Gan W, Reichhardt C~J~O and Lew W~S 2016 {\em Phys. Rev. B\/} {\bf 94}(14) 144405

\bibitem{vizarim_topological_2020}
Vizarim N~P, Reichhardt C, Reichhardt C~J~O and Venegas P~A 2020 {\em New Journal of Physics\/} {\bf 22} 053025

\bibitem{feilhauer_controlled_2020}
Feilhauer J, Saha S, Tobik J, Zelent M, Heyderman L~J and Mruczkiewicz M 2020 {\em Physical Review B\/} {\bf 102} 184425

\bibitem{gobel_skyrmion_2021}
G{\" o}bel B and Mertig I 2021 {\em Sci. Rep.\/} {\bf 11} 3020

\bibitem{reichhardt_magnus-induced_2015}
Reichhardt C, Ray D and Reichhardt C~J~O 2015 {\em New J. Phys.\/} {\bf 17} 073034

\bibitem{Yamaguchi20}
Yamaguchi R, Yamada K and Nakatani Y 2020 {\em Japan. J. Appl. Phys.\/} {\bf 60} 010904

\bibitem{souza_skyrmion_2021}
Souza J~C~B, Vizarim N~P, Reichhardt C~J~O, Reichhardt C and Venegas P~A 2021 {\em Phys. Rev. B\/} {\bf 104} 054434

\bibitem{purnama_guided_2015}
Purnama I, Gan W~L, Wong D~W and Lew W~S 2015 {\em Scientific Reports\/} {\bf 5} 10620

\bibitem{Reichhardt16a}
Reichhardt C and Reichhardt C~J~O 2016 {\em Phys. Rev. B\/} {\bf 94}(9) 094413

\bibitem{GonzalezGomez19}
Gonz\'alez-G\'omez L, Castell-Queralt J, Del-Valle N, Sanchez A and Navau C 2019 {\em Phys. Rev. B\/} {\bf 100}(5) 054440

\bibitem{Juge21}
Juge R, Bairagi K, Rana K~G, Vogel J, Sall M, Mailly D, Pham V~T, Zhang Q, Sisodia N, Foerster M, Aballe L, Belmeguenai M, Roussign{\' e} Y, Auffret S, Buda-Prejbeanu L~D, Gaudin G, Ravelosona D and Boulle O 2021 {\em Nano Lett.\/} {\bf 21}(7) 2989--2996

\bibitem{DelValle22}
Del-Valle N, Castell-Queralt J, Gonz{\' a}lez-G{\' o}mez L and Navau C 2022 {\em APL Mater.\/} {\bf 10} 010702

\bibitem{zhang_magnetic_2015}
Zhang X, Ezawa M and Zhou Y 2015 {\em Scientific Reports\/} {\bf 5} 9400

\bibitem{toscano_suppression_2020}
Toscano D, Mendonça J~P~A, Miranda A~L~S, Araujo C~I~L, Sato F, Coura P~Z and Leonel S~A 2020 {\em Journal of Magnetism and Magnetic Materials\/} {\bf 504} 166655

\bibitem{vizarim_soliton_2022}
Vizarim N~P, Souza J~C~B, Reichhardt C~J~O, Reichhardt C, Milošević M~V and Venegas P~A 2022 {\em Physical Review B\/} {\bf 105} 224409

\bibitem{vizarim_guided_2021}
Vizarim N~P, Reichhardt C, Venegas P~A and Reichhardt C~J~O 2021 {\em J. Mag. Mag. Mater.\/} {\bf 528} 167710

\bibitem{yanes_skyrmion_2019}
Yanes R, Garcia-Sanchez F, Luis R~F, Martinez E, Raposo V, Torres L and Lopez-Diaz L 2019 {\em Applied Physics Letters\/} {\bf 115} 132401

\bibitem{zhang_manipulation_2018}
Zhang S~L, Wang W~W, Burn D~M, Peng H, Berger H, Bauer A, Pfleiderer C, van~der Laan G and Hesjedal T 2018 {\em Nature Communications\/} {\bf 9} 2115

\bibitem{everschor_rotating_2012}
Everschor K, Garst M, Binz B, Jonietz F, Mühlbauer S, Pfleiderer C and Rosch A 2012 {\em Physical Review B\/} {\bf 86} 054432

\bibitem{kong_dynamics_2013}
Kong L and Zang J 2013 {\em Physical Review Letters\/} {\bf 111} 067203

\bibitem{zhang_skyrmion-skyrmion_2015}
Zhang X, Zhao G~P, Fangohr H, Liu J~P, Xia W~X, Xia J and Morvan F~J 2015 {\em Scientific Reports\/} {\bf 5} 7643

\bibitem{tomasello_strategy_2015}
Tomasello R, Martinez E, Zivieri R, Torres L, Carpentieri M and Finocchio G 2015 {\em Scientific Reports\/} {\bf 4} 6784

\bibitem{Feng22}
Feng Y, Zhang X, Zhao G and Xiang G 2022 {\em IEEE Trans. Electron Devices\/} {\bf 69} 1293

\bibitem{Jung21}
Jung D~H, Han H~S, Kim N, Kim G, Jeong S, Lee S, Kang M, Im M~Y and Lee K~S 2021 {\em Phys. Rev. B\/} {\bf 104}(6) L060408

\bibitem{souza_clogging_2022}
Souza J~C~B, Vizarim N~P, Reichhardt C~J~O, Reichhardt C and Venegas P~A 2022 {\em New J. Phys.\/} {\bf 24} 103030

\bibitem{evans_atomistic_2018}
Evans R~F~L 2018 Atomistic {Spin} {Dynamics} {\em Handbook of {Materials} {Modeling}: {Applications}: {Current} and {Emerging} {Materials}\/} ed Andreoni W and Yip S (Springer International Publishing) pp 1--23

\bibitem{iwasaki_universal_2013}
Iwasaki J, Mochizuki M and Nagaosa N 2013 {\em Nature Commun.\/} {\bf 4} 1463

\bibitem{iwasaki_current-induced_2013}
Iwasaki J, Mochizuki M and Nagaosa N 2013 {\em Nature Nanotechnology\/} {\bf 8} 742--747

\bibitem{seki_skyrmions_2016}
Seki S and Mochizuki M 2016 {\em Skyrmions in {Magnetic} {Materials}\/} {SpringerBriefs} in {Physics} (Springer International Publishing)

\bibitem{paul_role_2020}
Paul S, Haldar S, von Malottki S and Heinze S 2020 {\em Nature Communications\/} {\bf 11} 4756

\bibitem{gilbert_phenomenological_2004}
Gilbert T 2004 {\em {IEEE} Transactions on Magnetics\/} {\bf 40} 3443--3449

\bibitem{slonczewski_current-driven_1996}
Slonczewski J~C 1996 {\em Journal of Magnetism and Magnetic Materials\/} {\bf 159} L1--L7

\bibitem{zang_dynamics_2011}
Zang J, Mostovoy M, Han J~H and Nagaosa N 2011 {\em Physical Review Letters\/} {\bf 107} 136804

\bibitem{zhang_structural_2022}
Zhang X, Xia J and Liu X 2022 {\em Physical Review B\/} {\bf 105} 184402

\bibitem{zhang_particle-like_2022}
Zhang X, Xia J and Liu X 2022 {\em Physical Review B\/} {\bf 106} 094418

\bibitem{bellizotti_souza_spontaneous_2023}
Bellizotti~Souza J~C, Vizarim N~P, Reichhardt C~J~O, Reichhardt C and Venegas P~A 2023 {\em New Journal of Physics\/} {\bf 25} 053020 ISSN 1367-2630

\bibitem{kim_quantifying_2020}
Kim J~V and Mulkers J 2020 {\em IOP SciNotes\/} {\bf 1} 025211

\bibitem{bellizotti_souza_magnus_2022}
Bellizotti~Souza J, Vizarim N, Olson~Reichhardt C~J, Reichhardt C and Venegas P 2022 {\em Journal of Physics: Condensed Matter\/}

\bibitem{sampaio_nucleation_2013}
Sampaio J, Cros V, Rohart S, Thiaville A and Fert A 2013 {\em Nature Nanotechnology\/} {\bf 8} 839--844

\end{thebibliography}

\end{document}